\documentclass[amsmath,amssymb,aps,prb,twocolumn,superscriptaddress, reprint]{revtex4-2} 
\usepackage{svg}
\usepackage{graphicx}
\usepackage{dcolumn}
\usepackage{bm}
\usepackage{comment}
\usepackage{multirow}
\usepackage{array}
\usepackage{braket}
\usepackage{tabularx}
\usepackage{amsmath}
\usepackage{xcolor}
\usepackage[normalem]{ulem}
\usepackage{hyperref}
\usepackage{soul}
\setulcolor{red}
\usepackage{booktabs}
\usepackage{siunitx}
\DeclareUnicodeCharacter{2212}{\ensuremath{-}}

\makeatletter
\def\NAT@parse@name#1#2#3#4#5#6{%
  \ifcase#5\relax
    \NAT@parse@err
  \or
    \def\NAT@temp{{#1}{#2}{#3}{#4}}%
  \or
    \def\NAT@temp{{#1}{#2}{#3}{#4}}%
  \or  
    \def\NAT@temp{{#1}{#2}{#3}{#4}}%
  \or
    \def\NAT@temp{{#1}{#2}{#3}{#4}}%
  \else
    \def\NAT@temp{{#1}{#2}{#3}{#4}}%
    \def\NAT@temp@e{et al.}%
  \fi
  \expandafter\NAT@name\NAT@temp}
\makeatother

\begin{document}

\title{Beyond Germanides: Anomalous Hall Effect in the Silicide LaMn$_2$Si$_2$}

\author{Sergey V. Streltsov}
\email{streltsov@imp.uran.ru}
\affiliation{M. N. Mikheev Institute of Metal Physics, Ural Branch of Russian Academy of Sciences, 620137 Ekaterinburg, Russia}

\author{Dmitry M. Korotin}
\affiliation{M. N. Mikheev Institute of Metal Physics, Ural Branch of Russian Academy of Sciences, 620137 Ekaterinburg, Russia}

\date{\today}

\begin{abstract}
By combining symmetry analysis and direct density functional calculations including the spin-orbit coupling, we demonstrate  that anomalous Hall effect can be observed in not only germanides with general formula RMn$_2$Ge$_2$, where $R$ is a rare-earth ion or Y. Our calculations predict a large anomalous Hall conductivity in LaMn$_2$Si$_2$, with a non-zero $\sigma_{xy}^{AH}$ component of $-360~\text{S/cm}$, accompanied by a pronounced magneto-optical response. Remarkably, electron doping of LaMn$_2$Si$_2$ is expected to substantially enhance the Hall conductivity, with values reaching up to −650 S/cm. These results suggest that silicides with general formula RM$_2$Si$_2$ can be an interesting platform for studying anomalous Hall effect.
\end{abstract}

\maketitle

\section{Introduction}

Layered intermetallic compounds of the RMn$_2$Si$_2$ family, where R denotes a rare-earth element or yttrium, crystallize in the ThCr$_2$Si$_2$-type structure and have attracted considerable interest due to their complex magnetic ordering and relatively high magnetic transition temperatures~\cite{szytula1989, szytula1991,DiNapoli2004, DiNapoli2007, LaTbMn2Si2neutron, Mushnikov2012} ($T_c \sim 310 K$ for LaMn$_2$Si$_2$ and $T_c \sim 510 K$ for YMn$_2$Si$_2$). In this structure, Mn atoms form two-dimensional square lattices that stack along the crystallographic $c$-axis, separated by Si and R layers. The magnetism in these compounds is primarily governed by the Mn sublattice and can be finely tuned through chemical substitution or external pressure, yielding a diverse landscape of exchange-driven magnetic transitions~\cite{gerasimov2017, Gerasimov2018, Korotin2018, Korotin2020}.

\begin{figure}[b!]
\centering
\includegraphics[width=0.7\columnwidth]{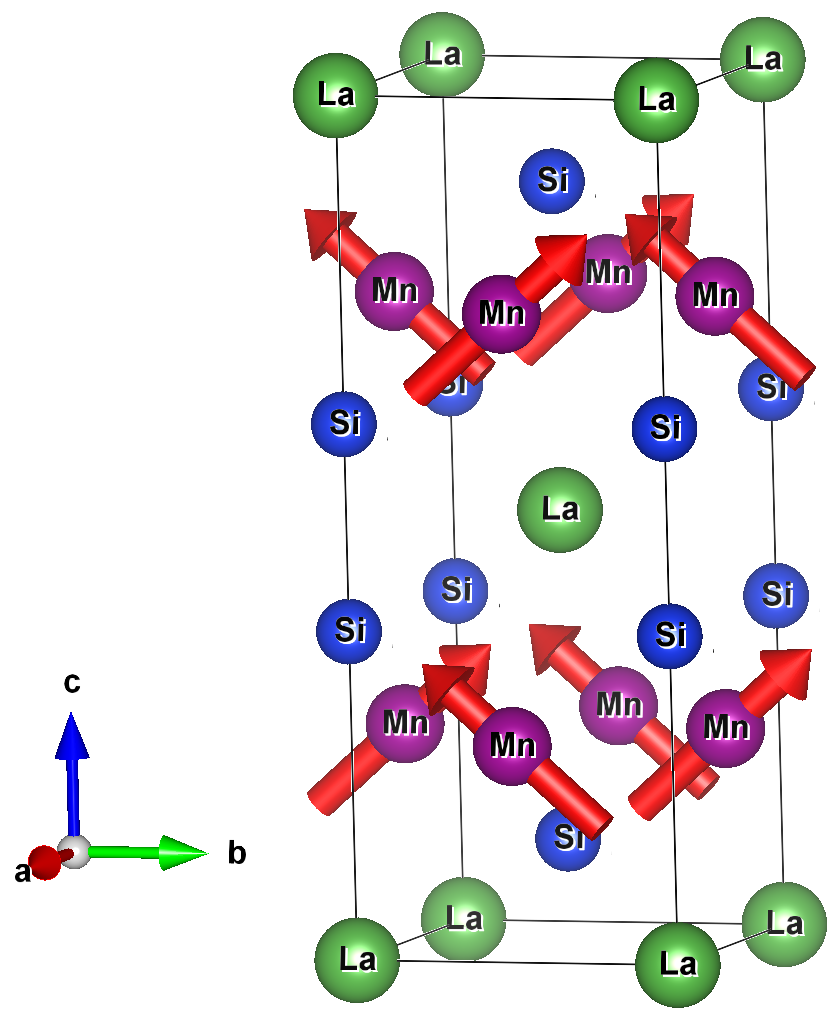}
    \caption{\label{fig:structure} Crystal and magnetic structure of LaMn$_2$Si$_2$.  Figure was plotted using VESTA~\cite{VESTA}.}
\end{figure}

The compound LaMn$_2$Si$_2$, which is the focus of this study, adopts a noncollinear magnetic structure, where Mn moments are ferromagnetically aligned along the $c$-axis while remaining antiferromagnetically coupled within the $ab$ plane ~\cite{Ijjaali1998, Gerasimov2018} (Fig.~\ref{fig:structure}). This canted spin arrangement breaks the inversion while preserving certain crystallographic symmetries. 

Recent studies on the isostructural compound LaMn$_2$Ge$_2$ have revealed giant topological Hall effect~\cite{Roychowdhury2024, Gong2021} at room temperature. This is related to a magnetic transition at 320K from antiferromagnetic to an incommensurate noncollinear structure with a ferromagnetic motif~\cite{Wang2022,Roychowdhury2024}. This motivates the investigation of physics related to Hall effect in LaMn$_2$Si$_2$ having a different magnetic structure. The noncollinear magnetic order in LaMn$_2$Si$_2$, characterized by the $Im'm2'$ magnetic space group. Our first-principles calculations with spin-orbit coupling demonstrate that this magnetic symmetry gives rise to momentum-dependent spin polarization of electronic bands and a substantial intrinsic anomalous Hall (AH) conductivity of $\sigma^{AH}_{xy}$. These results establish LaMn$_2$Si$_2$ as a prototype material for exploring transport in noncollinear ferromagnets, with implications for spintronic applications.

\section{Methods}
We used Vienna Ab initio Simulation Package (VASP) to calculate electronic and magnetic properties of LaMn$_2$Si$_2$. The generalized gradient approximation (GGA) in the form of Perdew-Burke-Ernzerhof exchange-correlation potential \cite{Perdew1997} and projected augmented wave (PAW)~\cite{Blochl1994} method were utilized \cite{Perdew1997}. 

The crystal structure was taken from Ref.~\cite{gerasimov2017}. The energy cutoff for the plane-wave basis was chosen to be 337 eV. We took into account the nonspherical contributions due to the gradient correction inside the PAW spheres. We used $12 \times 12 \times 7$-mesh in the reciprocal space for all summations. Although extrinsic effects from scattering can be substantial, this study is confined to the intrinsic anomalous Hall conductivity. This contribution To calculate the this contribution to the anomalous Hall conductivity tensor, we systematically increased the $k$-point mesh density until convergence was achieved. The final results, presented for a mesh size of 100$\times$100$\times$40, demonstrate satisfactory saturation of the conductivity values.

Previous DFT studies of RMn$_2$Si$_2$ demonstrated that one does not need to apply 
Hubbard $U$ correction for Mn $3d$ states, both electronic and magnetic properties can be described with out taking strong correlation effects into account~\cite{Korotin2017,Korotin2018,Korotin2020}. In the present paper the same strategy was adopted.

\section{Calculation results}
\subsection{Electronic and magnetic structure}

\begin{figure}[b!]
\centering
\includegraphics[width=1.1\columnwidth]{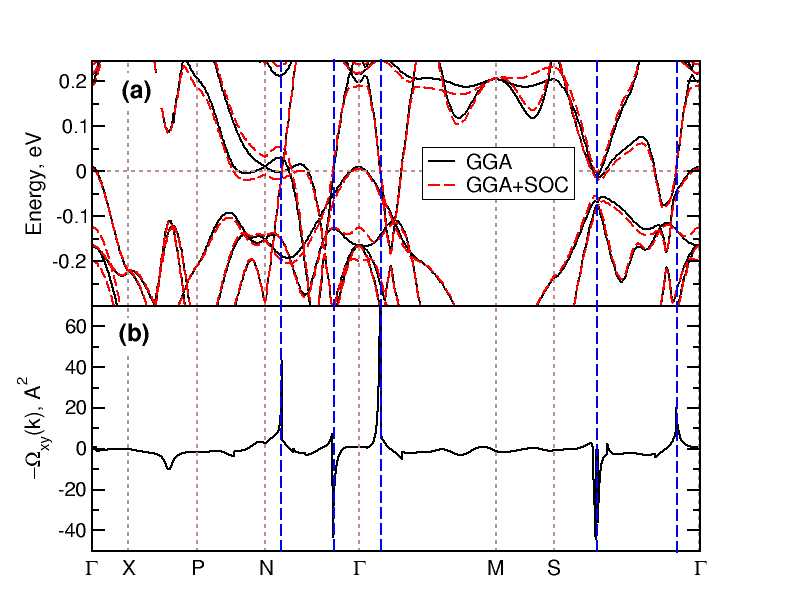}
    \caption{\label{fig:berry} Band structure as obtained in GGA and GGA+SOC calculations (a). Berry curvature $\Omega_{xy}$ along high-symmetry directions (b). The Fermi energy is at zero. Blue dashed lines show points along chosen path contributing the most to $\Omega_{xy} ({\bf k})$.}
\end{figure}

The non-collinear GGA calculation correctly reproduces the magnetic ground state of LaMn$_2$Si$_2$ refined in Ref. \cite{Gerasimov2018} and shown in Fig.~\ref{fig:structure}. The spin moments on Mn are  ${\bf m_{s}} = (\pm 2.55, 0, 1.55) \mu_B$, but they compensate each other in the $x$-direction in the spin space (such that the $x$-component of the total moment vanishes), while $m^z_s$ orders ferromagnetically. We note, that strictly speaking real and spin spaces are, of course, decoupled unless the spin-orbit coupling is introduced. The resulting band structure together with Berry curvature are presented in Fig.~\ref{fig:berry}. Interestingly, there are regions in the Brillouin zone where bands exhibit moderate dispersion — for example, near the $N$ point — along with several band crossings close to the Fermi level.

Including spin-orbit coupling (SOC) has almost no effect on the spin moments, but producing tiny orbital moments, ${\bf m_{l}} = (\pm 0.02, 0, 0.01) \mu_B$, but it changes the electronic structure resulting in formation of band anti-crossings, as one can see from Fig.~\ref{fig:berry}(a).

\begin{figure}[t!]
\centering
\includegraphics[width=0.72\columnwidth,angle=270]{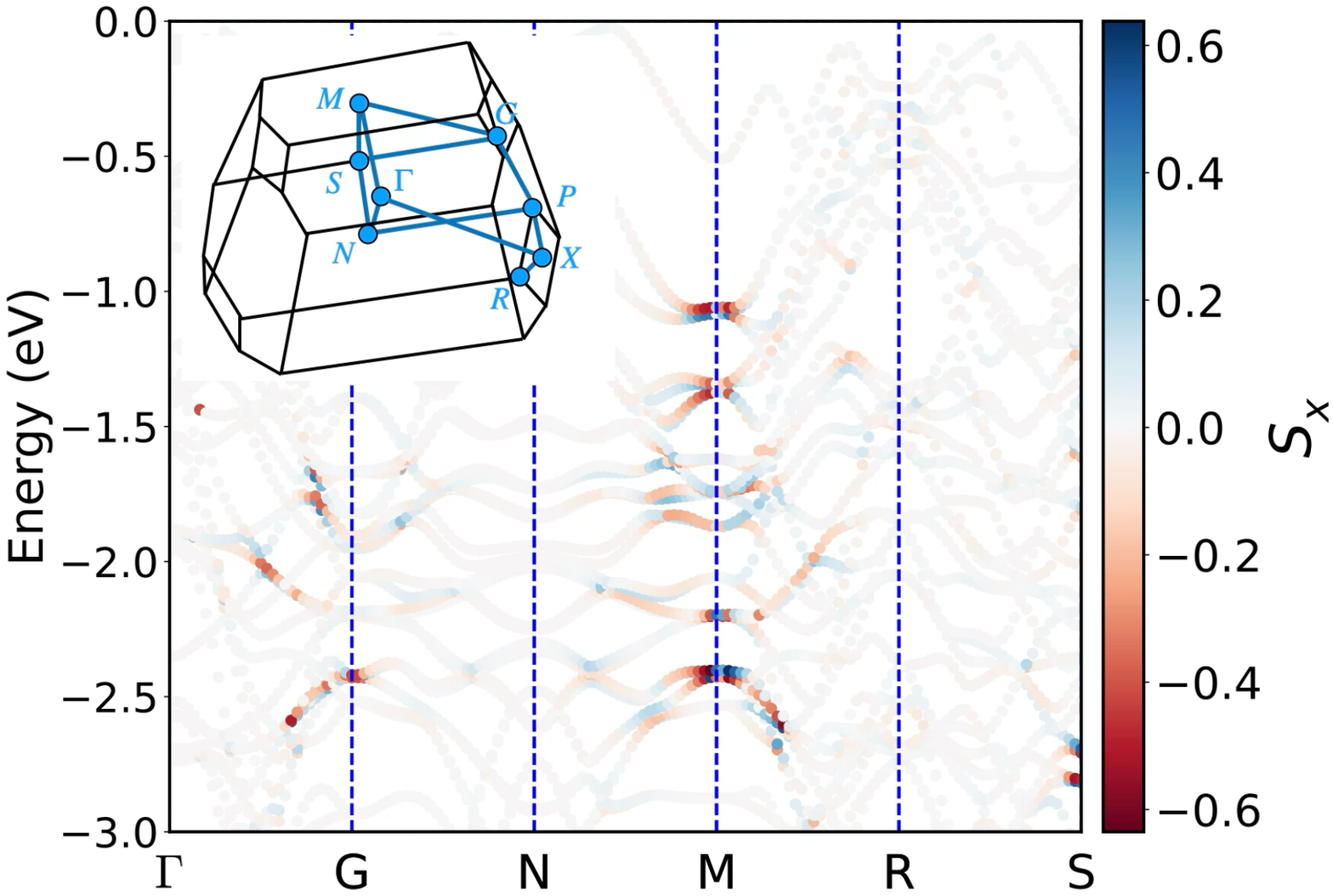}
    \caption{\label{fig:Sx-bands} GGA+SOC electronic band structure of LaMn$_2$Si$_2$ with $x$ projection of spin shown by color. The Fermi energy is set to zero. Inset shows k-points used for the band plotting.}
\end{figure}

Interestingly, although the magnetic structure of LaMn$_2$Si$_2$ (Fig.~\ref{fig:structure}) naturally provides a ferromagnetic component along the $z$-axis and vanishing magnetization along $x$, an analysis of the band structure reveals clear spin splitting in the valence band for the $S_x$ component near the G and M points (Fig.~\ref{fig:Sx-bands}).

\subsection{Symmetry analysis and anomalous Hall effect}
As it has been mentioned in the introduction the magnetic structure is characterized by the $Im'm2'$ magnetic space group. This magnetic order breaks the inversion center connecting two magnetic sublattices, the $C_4$ rotation axes, which make all Mn ions in the $ab$ plane equivalent, and one of the mirror planes present in the $I4/mmm$ space group. Finally, the magnetic point group turns out to be $m'm2'$. Here and below, we follow the conventional practice of considering magnetic symmetry using the same settings as for the parent paramagnetic structure. Therefore, we adopt non-standard settings (transformation to a standard for magnetic space group settings: $x \to x^s,\, y \to -z^s,\, z \to y^s $).

For the $m'm2'$ point group, $PT$ (parity $\times$ time-reversal) symmetry gets broken, and therefore one might expect various non-trivial physical phenomena. Formally, according to the classification of Cheong and Huang~\cite{cheong} this group belongs to the $M$-type of altermagnetism. However, its non-collinear antiferromagnetic structure may originate from exchange interaction rather than spin-orbit coupling. Therefore, classifying it as an altermagnet is premature. Nevertheless, the symmetry analysis performed using MTENSOR toolkit\cite{gallego2019} demonstrates that LaMn$_2$Si$_2$ should exhibit the anomalous Hall effect. In particular, the $xy$ component of conductivity is allowed to be non-zero, while all remaining off-diagonal elements must vanish. In order to verify this prediction we performed direct calculation of conductivity tensor $\sigma^{AH}_{\alpha\beta}$ by Wannier90 package\cite{Wannier90}, where it is estimated via the Kubo formula
\begin{eqnarray}
\label{Kubo}
\sigma^{AH}_{\alpha\beta} = -\frac {e^2}{\hbar} \frac 1 {N_k V_c} \sum_{{\bf k}}  \Omega_{\alpha\beta} ({\bf k}).
\end{eqnarray}
Here $\alpha$ and $\beta$ numerate Cartesian coordinates, $V_c$ is the cell volume, $N_k$ is the number of $k-$points, $\Omega_{\alpha\beta} ({\bf k}) = \sum_n f_{n {\bf k}} \Omega_{n, \alpha\beta} ({\bf k}) $ is total Berry curvature, $n,m$ are the band indexes, and $f_{n {\bf k}}$ is the Fermi-Dirac distribution.  $\Omega_{n, \alpha\beta} ({\bf k})$ can be evaluated as~\cite{wang2006a}
\begin{eqnarray}
\label{Kubo}
\Omega_{n, \alpha\beta} ({\bf k}) = -2 \hbar^2 {\rm Im}  \sum_{m \ne n} \frac{v_{nm,\alpha}({\bf k}) v_{mn,\beta}({\bf k})}{\left(\varepsilon_m({\bf k}) - \varepsilon_n({\bf k})\right)^2},
\end{eqnarray}
with band dispersion $\varepsilon_m({\bf k})$ and velocities $v_{nm,\alpha}({\bf k})$ obtained by wannierization.

We used Mn $3d$ and Si $3p$ states and a fine mesh in $k$-space for Wannier function projection. Integration in \eqref{Kubo} was performed over the 100$\times$100$\times$40 mesh and yielded $\sigma^{AH}_{xy} =  -365 \,{\rm S/cm}$, while both $\sigma^{AH}_{xz} $ and $\sigma^{AH}_{yz}$ are of order of few ${\rm S/cm}$, which perfectly agrees with symmetry consideration presented above.
\begin{figure}[t!]
\centering
\includegraphics[width=1\columnwidth]{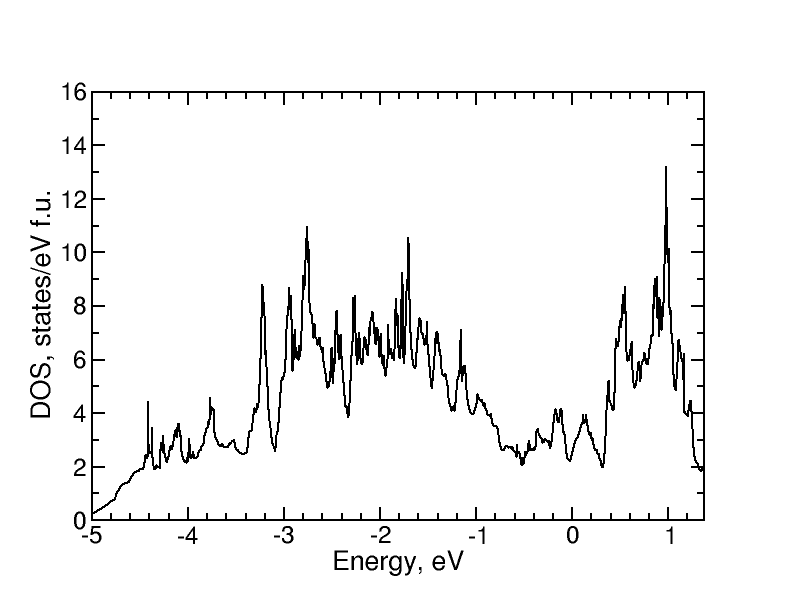}
    \caption{\label{fig:DOS}  Total density of states (DOS) as obtained in GGA+SOC calculations. The Fermi energy is set to zero.}
\end{figure}

Anomalous Hall conductivity in LaMn$_2$Si$_2$ is much higher (by absolute value) than large $\sigma^{AH}$ observed for non-collinear antiferromagnet Mn$_3$Sn (experimentally  $\sigma^{AH}_{exp} =  100 \,{\rm S/cm}$~\cite{nakatsuji2015}, while theoretical estimate is $\sigma^{AH}_{DFT} =  129 \,{\rm S/cm}$~\cite{Suzuki2017}), but it is still smaller when in germanide counterpart, LaMn$_2$Ge$_2$ , with $\sigma^{AH}_{exp} =928 \,{\rm S/cm}$~\cite{Gong2021,Roychowdhury2024}.

\begin{figure}[t!]
\centering
\includegraphics[width=1.1\columnwidth]{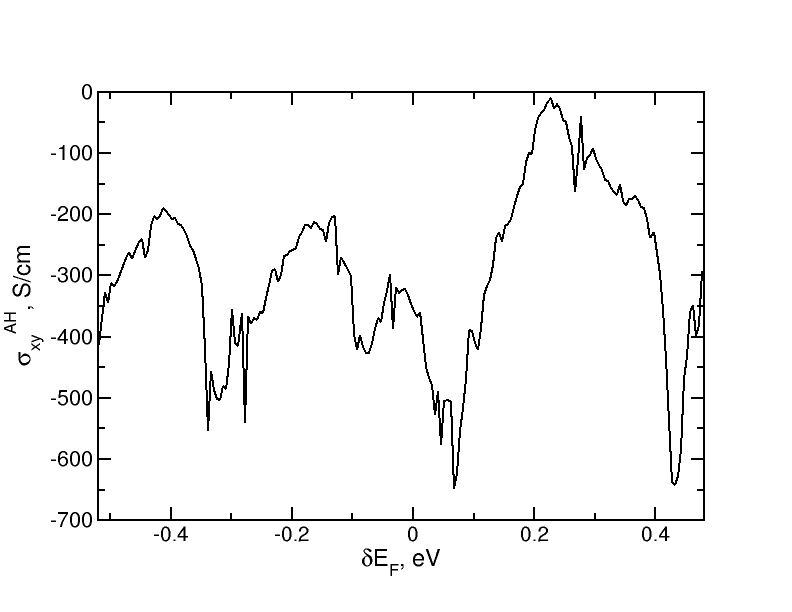}
    \caption{\label{fig:AHC} Intrinsic anomalous Hall conductivity $\sigma^{AH}_{xy}$ as function of the Fermi energy variation. Zero $\delta E_F$ corresponds to the actual Fermi level in LaMn$_2$Si$_2$.}
\end{figure}

Surprisingly, when treating LaMn$_2$Si$_2$ as a quasi-2D material containing a single unit cell (with the experimental $c = 10.56325$\AA \cite{LaTbMn2Si2neutron}), the calculated anomalous Hall conductivity is $\sigma^{AH}_{xy} = 0.996\,e^2/h$. This value is extremely close to the quantum conductance, which is rather unexpected for a metal and could potentially indicate the importance of topological features in the electronic structure. As one can see from Fig.~\ref{fig:DOS} there is no gap or pronounced pseudogap at the Fermi level. In Fig.~\ref{fig:AHC}, we plot the Hall conductance as a function of the Fermi energy. This analysis is useful both for studying doping effects and for demonstrating the absence of plateaus in the conductivity, suggesting that the observed $\sigma^{AH}_{xy} \approx e^2/h$ may be accidental.

We also note that a variation of $\delta E_F$ comparable with that presented in Fig~\ref{fig:AHC} does not alter the magnetic ground state.  The self-consistent GGA+SOC calculations show that even for $\delta E_F = 0.2$ (near the maximum of $\sigma^{AH}_{xy}$), the primary effect is only a slight decrease in the Mn moment canting to approximately $1.4^{\circ}$.

\begin{figure}[t!]
\includegraphics[width=1\columnwidth]{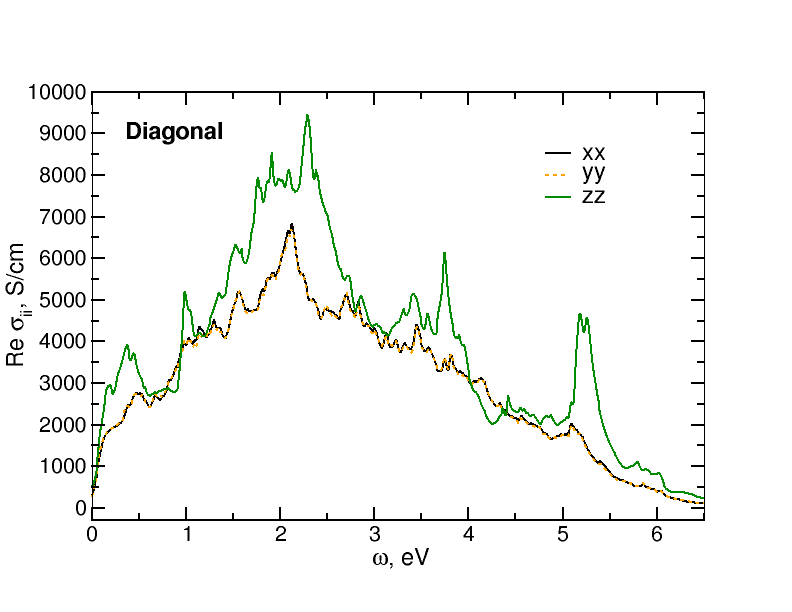}
\includegraphics[width=1\columnwidth]{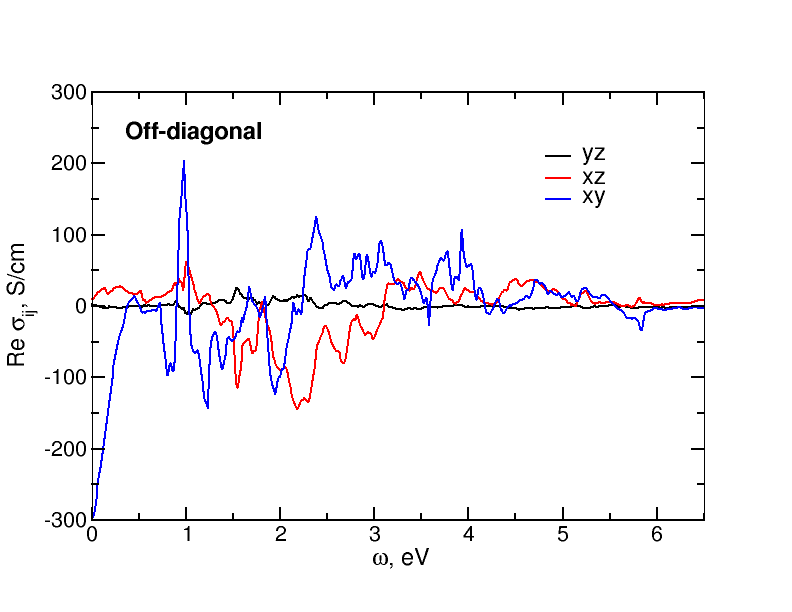}
 \caption{Diagonal (top panel) and off-diagonal (lower panel) components of optical conductivity $\sigma_ {\alpha\beta}$. One can see that the symmetry makes $\sigma_{yy} (\omega) = \sigma_{zz} (\omega)$. Off diagonal $\sigma_{xz}$ and $\sigma_{yz}$ vanish at zero frequency.
    \label{fig:sigma}}
\end{figure}

In Fig.~\ref{fig:sigma} we also present frequency dependence of diagonal and off-diagonal elements of optical conductivity calculated by Kubo formula. One might notice a low frequency feature of the $xy$ component, which not only saturates at static value corresponding to $\sigma^{AH}_{xy}$, but has a substantial contribution at optical frequencies.

\section{Discussions and conclusions}
Similar to anomalous Hall conductivity, one might expect spontaneous Nernst effect, which results in transverse electric field $E_i=\beta_{ij} \nabla_j T$ due to temperature gradient. For $N_{ij}=(\beta_{ij} - \pi_{ji})/2$ (where  $\pi_{ji}$ matrix relates the heat flux $q_i=\pi_{ij} J_j$ with current density, $J_j$), symmetry dictates that only $N_{xy} = N_{yx}$ can be finite.

Additional symmetry analysis demonstrates that there must be the direct and inverse piezomagnetic effects, similar to e.g. $M-$type altermagnet Ba$_3$NiRu$_2$O$_9$~\cite{streltsov2025b}. The direct effect is dependence of magnetization $M_i$ on the strain tensor $\sigma_{jk}$:
\begin{eqnarray}
M_i=\Lambda_{ijk}\sigma_{jk},    
\end{eqnarray}
where the piezomagnetic tensor $\Lambda_{ijk} \to \Lambda_{\alpha \beta}$ can be written using Voigt notations $xx \to 1, \; yy \to 2, \; zz \to 3, \; yz,zy \to 4, \; xz,zx \to 5, \; xy,yx \to 6$. In case of LaMn$_2$Si$_2$ 
\begin{eqnarray}
\Lambda_{\alpha \beta}=
\begin{pmatrix}
0 & 0 & 0 & 0 & \Lambda_{15} & 0 \\
0 & 0 & 0 & \Lambda_{24} & 0 & 0 \\
\Lambda_{31} & \Lambda_{32} & \Lambda_{33} & 0 & 0 & 0 
\end{pmatrix},
\end{eqnarray}
which results in the transverse piezomagnetic effect, when strain along $x$ or $y$ leads to magnetization in the $z$ direction, $M_z$.

To the best of our knowledge, neither the transverse piezomagnetic effect nor the anomalous Hall effect has been observed in RMn$_2$Si$_2$ materials so far. However, Hall conductivity measurements have been reported for their Ge counterparts such as CeMn$_2$(Ge$_x$Si$_{1-x})_2$ with $x=1$ and $x=0.2$~\cite{Levin1997},  LaMn$_2$Ge$_2$~\cite{Roychowdhury2024,Gong2021}, CeMn$_2$Ge$_2$\cite{Xu2022-1,Xu2022-2}, PrMn$_2$Ge$_2$\cite{Wang2023,Lyu2025}, NdMn$_2$Ge$_2$\cite{Zheng2021,Wang2020},  SmAg$_2$Ge$_2$~\cite{Bala2025}, and SmMn$_2$Ge$_2$~\cite{Huang2024,Singh2024}.

In the present paper, we show that not only germanides but also silicides with the ThCr$_2$Si$_2$ crystal structure can exhibit these phenomena. Moreover, as seen in Fig.~\ref{fig:AHC}, the absolute value of the Hall conductivity can be substantially increased by electron doping. Specifically, adding $\sim 0.4$ electrons per formula unit is expected to nearly double $\sigma_{xy}^{AH}$, provided the doping does not alter the magnetic structure — a scenario supported by the simplest GGA+SOC calculations.

Furthermore, a detailed symmetry analysis of various magnetic structures suggests that the anomalous Hall effect, and the transverse piezomagnetic effect may occur not only in LaMn$_2$Si$_2$ but also in many other ThCr$_2$Si$_2$-type materials where such phenomena have not yet been observed~\cite{bartashevich2025}. It would be highly compelling to investigate these effects in future experiments.

\section{Acknowledgments}
We thank S. Eremeev, A. Bartashevich, E.G. Gerasimov, A.F. Gubkin, and N.V. Mushnikov for stimulating discussions and  Ministry of Science and Higher Education of the Russian Federation for support, which came via Institute of Metal Physics.

\bibliography{Main2}

\end{document}